\documentclass[jcp,twocolumn,reprint]{revtex4-1}
\usepackage{amsfonts}
\usepackage{amsmath}
\usepackage{amssymb}
\usepackage{xcolor}
\usepackage{graphicx}%
\begin{document}
\title{Origin of the effective mobility in non-linear active micro-rheology}
\author{I. Santamar\'{\i}a-Holek}
\affiliation{UMJ-Facultad de Ciencias, 
Universidad Nacional Aut\'onoma de M\'exico, Campus Juriquilla, 
Boulevard Juriquilla, No. 3001, Juriquilla, C.P. 76230, Quer\'etaro, M\'exico}
\author{A. P\'{e}rez-Madrid}
\affiliation{Departament de F\'{\i}sica de la Mat\`{e}ria Condensada, Facultat de F\'{\i}sica, Universitat de
Barcelona, Mart\'{i} i Franqu\`{e}s 1, 08028 Barcelona, Spain}

\begin{abstract}
The distinction between the damping coefficient and the effective non-linear mobility of driven particles in active micro-rheology of supercooled liquids is explained in terms of individual and collective dynamics. The effective mobility arises as a collective effect which gives insight into the energy landscape of the system. On the other hand, the damping coefficient is a constant that modulates the  effect of external forces over the thermal energy which particles have at  their disposition to perform  Brownian motion.  For long times,  these thermal fluctuations become characterized in terms of an effective temperature that is a consequence of the dynamic coupling between kinetic and configurational degrees of freedom induced by the presence of the strong external force. The interplay between collective mobility and effective temperature allows to formulate a generalized Stokes-Einstein relation that may be used to determine the collective diffusion coefficient. The explicit relations we deduce reproduce simulation data remarkably well. 
\end{abstract}
\maketitle

%%%%%%%%%%%%%%%%%%%%%%%%%%%%%%%%%%%%%%%%%%%%%%%%%%%%%%%%%%%%%%%%%%%%%%%%%%%%%%%%%%%%%%%
\section{Introduction}
%%%%%%%%%%%%%%%%%%%%%%%%%%%%%%%%%%%%%%%%%%%%%%%%%%%%%%%%%%%%%%%x%%%%%%%%%%%%%%%%%%%%%%%%
It was recently  shown that a non-equilibrated energy transfer between kinetic and configuration degrees of freedom lies at the origin of a breakdown of the classical Stokes-Einstein relation in suspensions of particles at arbitrary concentrations \cite{Stokes2015}. This breakdown is quantified in terms of an effective temperature, $T_{eff}$, that depends on the local confining forces that a tracer particle experiences during its motion through the fluid \cite{JPCB2011}.  When a constant external force, $F$, is applied on the particles, the coupling between configuration and kinetic degrees of freedom is stronger and, consequently, the corresponding effective temperature contains the effect of this external force, $T_{eff}(F)$, \cite{JPCB2011}. A similar effect has been recently reported in simulations on active non-linear microrheology of supercooled liquids in Refs. \cite{binder,winter,heuer} in which the correspondence between simulation results and theory is remarkable, see, for instance, Ref. \cite{binder}.  It is convenient to notice that, although the methodology adopted in Ref.\cite{heuer}, where the authors use the potential energy landscape concept, differs notably from that of Refs. \cite{binder,winter}, in which correlation functions of dynamic variables are analyzed, the same effect associated with thermal fluctuations and the effective temperature appears. We will back to this point below.

If  $f(u,x,t)$ is the probability density in $(u,x)-$space, with $x$ and $u$ the position and velocity of a driven particle, then the heart of the statistical description of this problem is that for systems in or near an arrested state, like concentrated particle suspensions and supercooled liquids, the assumption of adiabatic decoupling 
\begin{equation}\label{probability}
f(u,x,t) \sim \phi_{eq}(u,t)\rho(x,t)
\end{equation}
{\it is no longer valid}, and therefore we have to use the exact definition
\begin{equation}\label{probability2}
f(u,x,t)\equiv \phi_{x}(u,t)\rho(x,t), %
\end{equation}
which means that the different time scales of the dynamics are not well separated. This simple fact has huge implications on the scaling of the effective temperature and mobility of glasses and glass like systems. In this work, we discuss these implications. In doing this, we provide a theoretical and conceptual framework that serves to understand, on the basis of a non-equilibrium description based on Fokker-Planck and Smoluchowski equations,  the origin of the already mentioned effective temperature and mobility, as well as to formulate a generalized version of the fluctuation-dissipation theorem in terms of the Stokes-Einstein relation, valid for far from equilibrium systems under the action of an external force. We illustrate our results by considering an ensemble of tracer particles in non-linear microrheology in atomic glasses.

The article is organized as follows. In Section II we explain the physics behind the non-adiabatic decoupling of the Fokker-Planck description. Then, InSection III we show the origin of the force dependent temperature whereas  Section IV is devoted to show the origin of the effective mobility within the mean field description. Finally, Section V discusses the generalization of the Stokes-Einstein relation and the Summary and Conclusions are presented in Section VI.

%%%%%%%%%%%%%%%%%%%%%%%%%%%%%%%%%%%%%%%%%%%%%%%%%%%%%%%%%%%%%%%x%%%%%%%%%%%%%%%%%%%%%%%%
\section{Non adiabatic decoupling and Fokker-Planck equation}
%%%%%%%%%%%%%%%%%%%%%%%%%%%%%%%%%%%%%%%%%%%%%%%%%%%%%%%%%%%%%%%x%%%%%%%%%%%%%%%%%%%%%%%%

In order to proceed, it is convenient to introduce  the well-known Kramers-Fokker-Planck equation
\begin{equation}\label{Fokker-Planck}
\frac{\partial}{\partial t}f=-u\nabla f+\nabla
V(x)\frac{\partial}{\partial u}f
+\beta \frac{\partial}{\partial u} \left(  \frac{k_{B}T_{0}}{m}\frac{\partial}{\partial
u}f+uf\right)  %
\end{equation}
which describes the dynamics of an ensemble of non-interacting Brownian degrees of freedom, that is, the velocity $u$ and the position $x$ of the driven particles in contact with a heat bath. In Eq. (\ref{Fokker-Planck}) $V(x)$ stands for the specific energy landscape, $m$ the mass of the driven particles, $T_0$ is the temperature of the heat bath and $k_B$ the Boltzmann's constant. It is convenient to stress that $\beta$ is known as the damping  coefficient (friction per unit of mass) of a single particle which may differ, in the general case, from the effective friction coefficient $\beta_{eff}=1/\mu$ of the ensemble, with $\mu$ the mobility.

Our aim is to calculate the explicit expression of $\beta_{eff}(F)$. Usually, this is done starting from the constrained description in terms of the reduced density $\rho(x,t)= \int f(u,x,t)du$.\cite{JPCB2011}  Note here that, the normalization of the probability is expressed by 
\begin{equation}\label{normalization}
\int f(u,x,t)dxdu=1.
\end{equation}
The configurational probability density evolves according to%
\begin{equation}\label{configurational probability}
\frac{\partial}{\partial t}\rho(x,t)=-\nabla \left[\int uf(u,x,t)du\right]. %
\end{equation}
Equation (\ref{configurational probability}) has been obtained by partial integration of Eq. (\ref{Fokker-Planck})
over velocity and thus, implicitly defines the current $J(x,t)= \int
uf(u,x,t)du$. For times $t>>\beta^{-1}$, the constitutive relation
for $J(x,t)$ can be obtained by taking the time derivative of 
$\int uf(u,x,t)du$, using Eq. (\ref{Fokker-Planck}) and integrating by parts assuming that the
currents vanish at the boundaries. Thus, after neglecting the time derivative 
$\partial J(x,t)/\partial t$, the resulting expression is 
\begin{equation}
\label{constituttive}
J(x,t)=- \beta^{-1} \nabla \left[\left(\int u^{2} \phi_x(u)du\right) \, \rho\right]
 - \beta^{-1} \rho \nabla V.
\end{equation}
Large external forces cause local inhomogeneities in the system in
such a way that there is no clear time separation between the relaxation of
the fast and the slow (configuration) variables, a situation which corresponds
to the underdamped motion of the driven particles.

In these conditions, $\phi_{x}(u,t)$ is not given by a Maxwellian and thus, according to the equipartition theorem, the second
moment of $\phi_{x}(u,t)$ yields a nonequilibrium  temperature
$T(x,t)$
\begin{equation}
\frac{k_{B}T(x,t)}{m}=\int u^{2}\phi_{x}(u,t)du. \label{nonequilibrium temperature}%
\end{equation}
% Note that the kinetic temperature introduced through Eq. (\ref{nonequilibrium temperature}) is consistent with the familiar expression
%\begin{equation} \label{nonequilibrium temperature2}
%k_{B}T(x,t)=\dfrac{1}{m^2}\int p^2\hat{\phi}_{x}(p,t)dp,
%\end{equation}
%where $\hat{\phi}_{x}(p,t)\equiv\phi_{x}(u(p),t)$.

Since it is not possible to know a \textit{priori} the explicit dependencies of $\phi_{x}(u,t)$, it is convenient to make an overall estimate of this effective temperature.
Defining $\phi(u,t) \equiv  \int \phi_{x}(u,t) \rho(x,t) dx$ and using Eq. (\ref{Fokker-Planck}), we obtain the modified Rayleigh equation 
\begin{equation}\label{Rayleigh}
\frac{\partial \phi}{\partial t}=\beta\frac{\partial}{\partial
u}\left[  u\phi+\frac{k_{B}T_{0}}{m}\frac{\partial \phi}{\partial u}%
\right]  +\frac{\partial}{\partial u}\left[  \phi \langle\nabla V\rangle\right]  ,
\end{equation}
where we have introduced the volume averaged specific force: 
$\langle\nabla V\rangle = \int \rho(x,t) \nabla V(x)  dx$. The Rayleigh equation (\ref{Rayleigh})  is valid up to zero order in the expansion
 \begin{equation} \label{expansion}
 \phi_{x}(u,t) \simeq \phi(u,t)+ \frac{\nabla V}{L \beta^{2}}\delta\phi(u,t)+O\left( \nabla V^{2}\right).
 \end{equation}
where $L$ is a characteristic length of the system. Eq. (\ref{Rayleigh}) describes the relaxation of the spatially-averaged distribution $\phi$ and, as a consequence of this, it includes, over average, the effect of the external forces on the dynamics of the fast variable through the \emph{time dependent
quantity} $\langle\nabla V(t)\rangle$.

%%%%%%%%%%%%%%%%%%%%%%%%%%%%%%%%%%%%%%%%%%%%%%%%%%%%%%%%%%%%%%%x%%%%%%%%%%%%%%%%%%%%%%%%
\section{Origin of the force dependent effective temperature}
%%%%%%%%%%%%%%%%%%%%%%%%%%%%%%%%%%%%%%%%%%%%%%%%%%%%%%%%%%%%%%%x%%%%%%%%%%%%%%%%%%%%%%%%

The solution of Eq. (\ref{Rayleigh}) can be found by using the method of characteristics. Thus, in the absence of diffusion, the velocity $u$ is determined by the initial velocity $u_{0}$ through the relation
\begin{equation}\label{initial velocity}
u=u_{0}e^{-\beta t}-\int_{0}^{t}e^{-\beta\left(  t-s\right)  }\langle\nabla V(s)\rangle ds.
\end{equation}
%where $f(t)=\langle\nabla V(t)\rangle$. 
A change of variables from $u$ to
\begin{equation}\label{initial velocity2}
y=u e^{\beta t}+\int_{0}^{t}e^{\beta s}\langle\nabla V(s)\rangle ds
\end{equation}
transforms Eq. (\ref{Rayleigh}) into
\begin{equation}\label{Rayleigh2}
\frac{\partial}{\partial t}\phi(y,t)=\beta\phi(y,t)+\frac
{k_{B}T_{0}}{m}\beta e^{2\beta t}\frac{\partial^{2}}{\partial y^{2}}\phi(y,t).
\end{equation}
A second change, $\psi=$ $e^{-\beta t}\phi$ leads to%
\begin{equation}\label{Rayleigh3}
\frac{\partial}{\partial t}\psi=\frac{k_{B}T_{0}}{m}\beta
 e^{2\beta t}\frac{\partial^{2}}{\partial y^{2}}\psi, %
\end{equation}
whose solution is a Gaussian\cite{Chandrasekhar} that, written in the original variables, reads
%\begin{equation}
%\psi=\frac{1}{\sqrt{4\pi\int_{0}^{t}\triangle^{2}(t^{\prime})dt^{\prime}}}%
%\exp\left[  -(y-y_{0})^{2}/4\int_{0}^{t}\triangle^{2}(t^{\prime})dt^{\prime
%}\right]  ,\label{a4}%
%\end{equation}
%with $\triangle^{2}(t)=(k_{B}T_{0}/m)\beta e^{2\beta t}$
%and $y_{0}=u_{0}$. Hence,
\begin{equation}\label{Gaussian}
\phi=\frac{1}{\sqrt{\frac{2\pi k_{B}T_{0}}{m}\left(  1-e^{-2\beta
t}\right)  }}\times
\end{equation}
\begin{equation}\nonumber
\exp\left\{  -\frac{\left[  u-e^{-\beta t}\left(
u_{0}-\int_{0}^{t}e^{\beta s}\langle\nabla V(s)\rangle ds\right)
\right]  ^{2}}{\frac{2k_{B}T_{0}}{m}\left(  1-e^{-2\beta t}\right)
}\right\}  
\end{equation}
At long time the dependence of $\phi$ on $u_{0}$ disappears, and if $\langle\nabla V(t)\rangle$ is a slow varying function of is argument such that%
\begin{gather}\label{Gaussian2}
\int_{0}^{t}e^{-\beta\left(  t-s\right)  }\langle\nabla
V(s)\rangle ds=\int_{0}^{t}e^{-\beta s}\langle\nabla V(t-s)\rangle
ds\\\nonumber
\simeq\langle\nabla V(t)\rangle\int_{0}^{\infty}e^{-\beta s}%
ds=\langle\nabla V(t)\rangle\beta^{-1},\label{a6}%
\end{gather}
then the long time \textit{quasi-stationary} limit of Eq. (\ref{Gaussian})  becomes%
\begin{equation}\label{Gaussian3}
\phi=\frac{1}{\sqrt{\frac{2\pi k_{B}T_{0}}{m}}}\exp\left\{  -\frac{\left[
u-\langle\nabla V(t)\rangle/\beta\right]  ^{2}}{\frac{2k_{B}T_{0}%
}{m}}\right\}.  
\end{equation}
The second moment of Eq. (\ref{Gaussian3}) is then given by
\begin{equation}\label{second moment}
\overline{u^{2}}=\frac{k_{B}T_{0}}{m}+\left[\langle\nabla V(t)\rangle/\beta\right]^{2}\equiv \frac{k_{B}T_{eff}(t)}{m}.
\end{equation}
This equation is the spatial average of the nonequilibrium temperature $T(x,t)$ introduced in Eq. (\ref{nonequilibrium temperature}): $T_{eff} \simeq \langle T(x,t) \rangle$. It is worth emphasizing that Eq. (\ref{second moment}) contains the damping coefficient $\beta$ in a form that weighs the corrective effect that external forces induce on the thermal energy that particles have at their  disposal in order to perform  Brownian motion. This extra energy comes from the non-equilibrated energy distribution associated to velocities.\cite{Stokes2015} Thus, Eq. (\ref{second moment}) can be interpreted as an equipartition theorem defining the thermal energy of the system that is proportional to the non-equilibrium mean-field effective temperature $T_{eff}$. 
When the specific potential energy is of the form $V(x)=U(x)-Fx$, due to the application of an external force $F$,  Eq. (\ref{second moment}) reduces to
\begin{equation}\label{meanfield temperature}
T_{eff}(F) =T_{0}+\frac{m}{k_B\beta^{2}} \left(  \langle\nabla U(x)\rangle- F\right)^{2} .
\end{equation}
In the literature reporting simulations on nonlinear-active microrheology, different strategies were used to introduce the effective temperature similar to Eq. (\ref{meanfield temperature}). For instance in Refs. \cite{binder,winter}, the authors inferred its existence by analyzing the behavior of the mean square displacement and the diffusion coefficient in terms of the externally applied force. More recently, in Ref. \cite{heuer} the potential energy landscape approach was used in order to calculate the population distribution of different metabasin energies in the presence of the external force. It was shown that the energy distribution of the system maintains a Gaussian form but becomes shifted towards higher energies, fact that
yield the authors to characterize the effect in terms of an effective temperature whose dependence on the force coincides with Eq. (\ref{meanfield temperature}). 

%%%%%%%%%%%%%%%%%%%%%%%%%%%%%%%%%%%%%%%%%%%%%%%%%%%%%%%%%%%%%%%x%%%%%%%%%%%%%%%%%%%%%%%%
\section{Origin of the force dependent effective mobility: The mean field description}
%%%%%%%%%%%%%%%%%%%%%%%%%%%%%%%%%%%%%%%%%%%%%%%%%%%%%%%%%%%%%%%x%%%%%%%%%%%%%%%%%%%%%%%%

Adopting now a mean-field description by using $T_{eff}(F)$ in the first term at the right hand side of Eq. (\ref{constituttive}) yields 
%the following expression for the diffusion current
\begin{equation}\label{diffusion current}
J(x,t) \simeq - \frac{k_{B}T_{eff}}{m\beta}  \nabla\rho(x,t) - \rho(x,t) \left[ \nabla U(x) - F\right]  , %
\end{equation}
which, after substituted into Eq. (\ref{configurational probability}), yields the
\emph{mean-field Smoluchowski equation}
\begin{equation}\label{Smoluchowski}
\frac{\partial}{\partial t}\rho=D_{eff}\nabla^{2}\rho-\frac{1}{\beta}\nabla\left\{  \rho\left[ \nabla U(x) - F\right]\right\} ,
\end{equation}
where we have introduced the mean-field (self) diffusion coefficient 
\begin{equation}\label{diffusion coefficient}
D_{mf}=k_{B}T_{eff} \mu_0,
\end{equation}
where we have introduced the bare mobility coefficient $\mu_0= ({m\beta})^{-1}$ that depends on the damping coefficient $\beta$.

The calculation of the effective  mobility coefficient can now be done as follows. First, we define the local velocity $v(x,t)$ as
\begin{equation}\label{local velocity}
J(x,t)\equiv\varrho(x,t)v(x,t)
\end{equation}
such that $\left\langle J(x,t)\right\rangle =\left\langle v(x,t)\right\rangle$ and  according to Eq. (\ref{diffusion current}), see Refs. \cite{Kenkre,Stratonovich},
\begin{equation} \label{definition mu}
\langle v(t) \rangle 
= \frac{1}{\beta} \int_0^L { \rho(x,t)\left[\nabla U(x) - F\right]}dx ,%
\end{equation}
where the characteristic length of a cage $L$, coincides with the periodicity of the potential energy $U(x)=U(x+L)$. In first approximation, we may substitute the stationary solution $\rho_{st}(x)$ of Eq. (\ref{Smoluchowski}) in order to obtain the following expression for the effective mobility $\mu(F)$:  
\begin{equation} \label{effective mobility}
\mu(F)= -\frac{\langle v(t) \rangle}{F} = \mu_0 \frac{k_BT_{eff}}{F}\frac{\left(1-e^{-FL/k_BT_{eff}}\right) }{A_{L,F} }.%
\end{equation}
The factor $A_{L,F}$ in Eq. (\ref{effective mobility}) is defined by
\begin{equation} \label{A}
A_{L,F}= \frac{1}{L} \int_0^L e^{-Fx/k_BT_{eff}}  \int_0^L e^{U(z)-U(z+x)/k_BT_{eff}}dz dx ,%
\end{equation}
and has dimensions of length. In the limit $L \rightarrow \infty$, it has been shown that $\mu(F) = \beta^{-1}$. 
\begin{figure}
{} \mbox{\resizebox*{8.0cm}{5.0cm}{\includegraphics{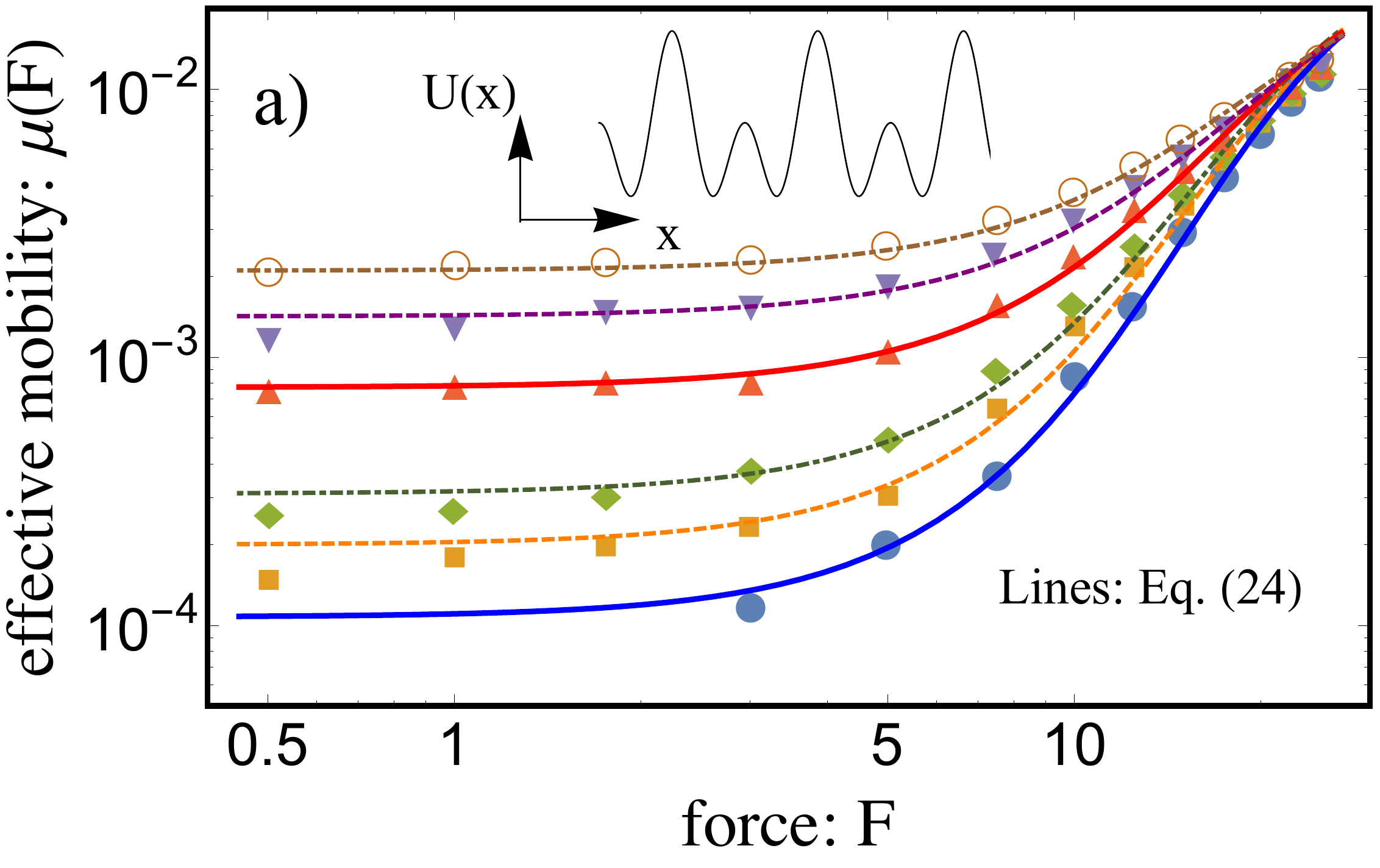}}}
{} \mbox{\resizebox*{8.0cm}{5.0cm}{\includegraphics{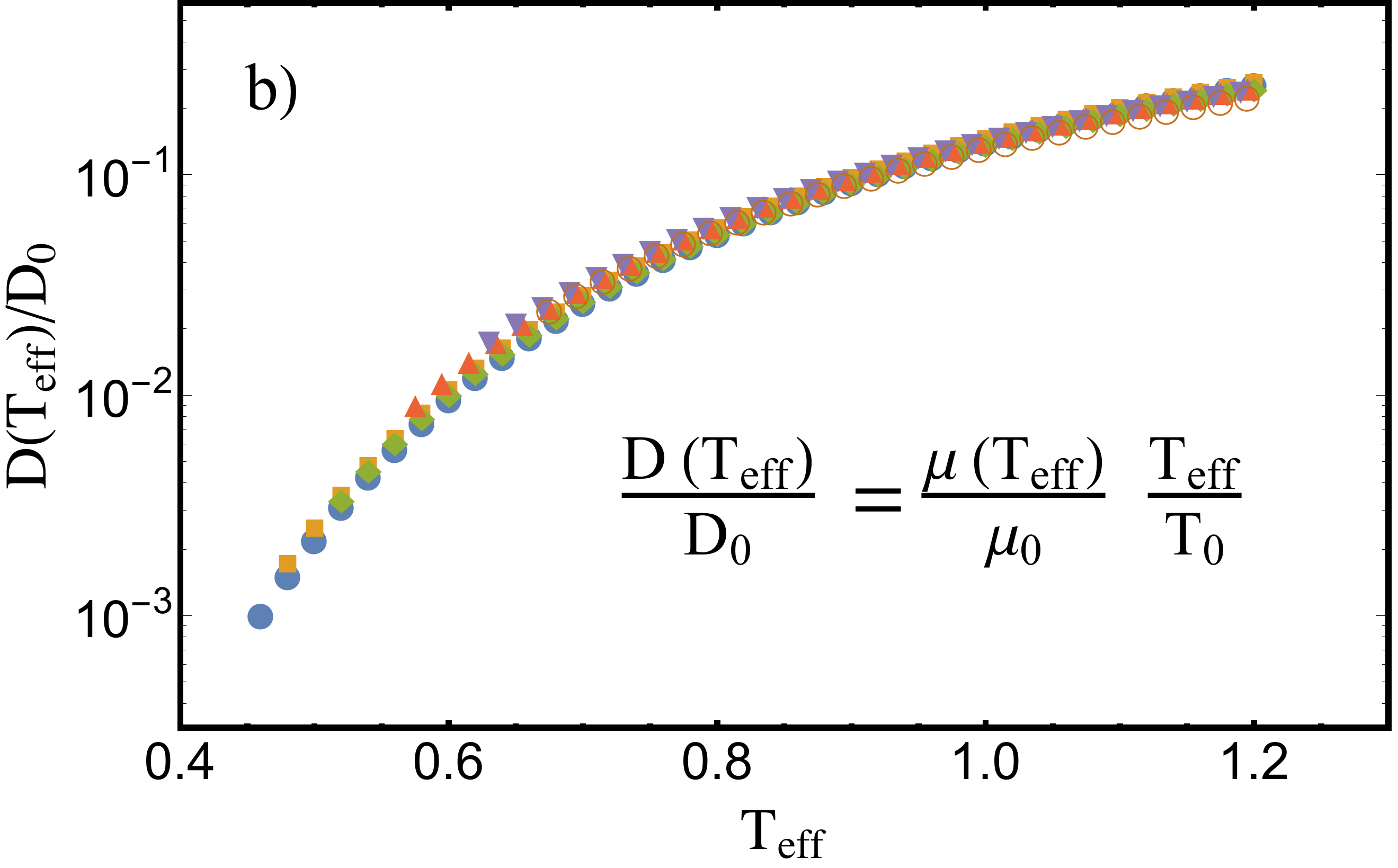}}}
\caption{a) Comparison between simulation data (symbols) taken from Ref. \cite{heuer} and the results for the effective mobility given by Eq. (\ref{effective mobility}) (lines) assuming the periodic potential $U(x)$ given in Eq. (\ref{periodic potential}). b) Normalized  effective diffusion coefficient as a function of the effective temperature according to Eq. (\ref{GSER}). The symbol's colors correspond to those of a). } 
\end{figure}

Equation (\ref{effective mobility}) shows that the effective mobility $\mu$ is a non-linear function of the external force $F$, no matter if the mean field temperature $T_{eff}$ is or not a function of $F$.  Assuming that the periodic potential energy is of the form
\begin{equation} \label{periodic potential}
U(x)= Uo\,\left(\cos^2\left[\frac{2\pi x}{L}\right]+\frac{8}{10}\sin^2\left[\frac{2\pi x}{L}\right]\right)%
\end{equation}
and approximating the arguments of the exponentials in Eq. (\ref{A}) by their first order series expansion in terms of $z$, it is possible to compare our results with the simulation data for the dependence of the mobility coefficient on the externally applied force \cite{heuer}, see Fig. 1a. Other potentials may be used to make the interpolation of simulation data (not shown). The set of parameters change but the fit is equally good. We assumed the potential (\ref{periodic potential}), shown in Fig. 1a, because it has a combination of large and small energy barriers. It is convenient to mention that the potential energy $U(x)$ is an effective potential, a mean force potential, whose parameters may depend upon the temperature \cite{HillLibro}. For the interpolation of the simulation data in in Fig. 1a we have assumed $m=1$ and $L=1$ in all cases. For the blue circles (data) and the blue solid line we have used $T=0.450$, $\mu_0 \simeq 0.0353$ and $U_0 \simeq 0.6026$; for the orange squares and the orange dashed line we have used $T=0.475$, $\mu_0 \simeq 0.0351$ and $U_0 \simeq 0.5918$; for the green diamonds and the green dotted-dashed line we have used $T=0.500$, $\mu_0 \simeq 0.0348$ and $U_0 \simeq 0.5891$; for the red upward triangles and the red solid line we have used $T=0.550$, $\mu_0 \simeq 0.0331$ and $U_0 \simeq 0.5649$, for the purple downward triangles and the purple dashed line we have used $T=0.600$, $\mu_0 \simeq 0.0316$ and $U_0 \simeq 0.5515$ and, finally, for the purple downward triangles and the purple dashed line we have used $T=0.650$, $\mu_0 \simeq 0.0311$ and $U_0 \simeq 0.5515$. Units are arbitrary.

%%%%%%%%%%%%%%%%%%%%%%%%%%%%%%%%%%%%%%%%%%%%%%%%%%%%%%%%%%%%%%%x%%%%%%%%%%%%%%%%%%%%%%%%
\section{The generalized Stokes-Einstein relation}
%%%%%%%%%%%%%%%%%%%%%%%%%%%%%%%%%%%%%%%%%%%%%%%%%%%%%%%%%%%%%%%x%%%%%%%%%%%%%%%%%%%%%%%%
An interesting question regarding these results is related with the fluctuation-dissipation theorem. In Ref. \cite{Stokes2015}, it was shown for colloidal suspensions at arbitrary volume fractions that the Stokes-Einstein relation has to be modified by considering the natural scaling of all the quantities involved, the effective diffusion and viscosity coefficients and the effective temperature, in order to cope with experimental data. Following this idea, we propose that for molecular glasses a generalized Stokes-Einstein relation similar to Eq. (3) of Ref. \cite{Stokes2015} may hold, that is:
\begin{equation}\label{GSER}
\frac{D(T_{eff})}{D_0} = \frac{\mu(T_{eff})}{\mu_0}\,\frac{T_{eff}}{T_0},
\end{equation}
where we have introduced the long-time collective diffusion coefficient $D(T_{eff})$. Fig 1b shows the  scaling of this quantity in terms of the effective temperature $T_{eff}$. This result means that the effective temperature is the natural scaling variable of the system, in similar form as the effective volume fraction is the natural scaling variable of concentrated colloidal suspensions \cite{Stokes2015}. Other calculations showing correction of the diffusion coefficient when Brownian particles move along periodic and tilted potentials are reported in Ref. \cite{AgustinPRL2001}.

%%%%%%%%%%%%%%%%%%%%%%%%%%%%%%%%%%%%%%%%%%%%%%%%%%%%%%%%%%%%%%%x%%%%%%%%%%%%%%%%%%%%%%%%
\section{Summary and conclusions}
%%%%%%%%%%%%%%%%%%%%%%%%%%%%%%%%%%%%%%%%%%%%%%%%%%%%%%%%%%%%%%%x%%%%%%%%%%%%%%%%%%%%%%%%

As a summary, in this work we have shown the distinction between the damping coefficient and the effective non-linear mobility of driven particles in active micro-rheology of supercooled liquids. In addition, we have shown that the origins of the effective temperature and of the effective mobility also differ. The effective temperature is a consequence of the dynamic coupling between kinetic and configurational degrees of freedom. The effective mobility arises as a collective effect that gives insight into the energy landscape of the system. From the interplay between these quantities it was possible to formulate a generalized Stokes-Einstein relation that may be used to determine the collective diffusion coefficient in non-linear microrheology of atomic glasses. Comparison of our theoretical results with simulation data is remarkably good. We believe that our findings may be of utility in understanding the thermo-dynamics of glassy systems.

%%%%%%%%%%%%%%%%%%%%%%%%%%%%%%%%%%%%%%%%%%%%%%%%%%%%%%%%%%%%%%%x%%%%%%%%%%%%%%%%%%%%%%%%
\section*{Acknowledgements}
%%%%%%%%%%%%%%%%%%%%%%%%%%%%%%%%%%%%%%%%%%%%%%%%%%%%%%%%%%%%%%%x%%%%%%%%%%%%%%%%%%%%%%%%
This work was supported by the DGiCYT of Spanish
Government under Grant No. FIS2015-67837-P and by UNAM DGAPA Grant No.
IN113415. We also thank the Academic mobility program between the University
of Barcelona and the National Autonomous University of Mexico.

\end{document}